\documentclass[12pt,a4paper]{article}
\usepackage{a4wide}
\usepackage{amsmath}
\usepackage{amssymb}
\usepackage{epsfig}
\usepackage{subfigure}
\usepackage{exscale}
\usepackage{float}
\usepackage{bbm}
\usepackage[numbers,sort&compress]{natbib}
\usepackage{pst-plot, pstricks,pst-math}
\usepackage{fancybox,amssymb,color}
\usepackage{graphicx}
\usepackage{pstricks, color, graphicx, epsfig, psfrag}
\usepackage{amsfonts,amsmath,amssymb,slashed}
\usepackage{dsfont}
\usepackage{bbm,bm}
\usepackage{fancyhdr, a4wide}
\usepackage[english]{babel}
\usepackage{subfigure}

\makeatletter
\@addtoreset{equation}{section}
\makeatother

\title{Low-Temperature Properties of Two-Dimensional Ideal Ferromagnets}

\author{Christoph P.\ Hofmann$^a$ \\ \\
\normalsize {$^a$ Facultad de Ciencias, Universidad de Colima} \\
\vspace{0.3cm}
\normalsize {Bernal D\'iaz del Castillo 340, Colima C.P.\ 28045, Mexico} \\}

\begin{document}

\maketitle

\begin{abstract} \normalsize

\end{abstract}

The manifestation of the spin-wave interaction in the low-temperature series of the partition function has been investigated extensively
over more than seven decades in the case of the three-dimensional ferromagnet. Surprisingly, the same problem regarding ferromagnets in
two spatial dimensions, to the best of our knowledge, has never been addressed in a systematic way so far. In the present paper the
low-temperature properties of two-dimensional ideal ferromagnets are analyzed within the model-independent method of effective
Lagrangians. The low-temperature expansion of the partition function is evaluated up to two-loop order and the general structure of this
series is discussed, including the effect of a weak external magnetic field. Our results apply to two-dimensional ideal ferromagnets which
exhibit a spontaneously broken spin rotation symmetry O(3) $\to $ O(2) and are defined on a square, honeycomb, triangular or Kagom\'e
lattice. Remarkably, the spin-wave interaction only sets in at three-loop order. In particular, there is no interaction term of order
$T^3$ in the low-temperature series for the free energy density. This is the analog of the statement that, in the case of
three-dimensional ferromagnets, there is no interaction term of order $T^4$ in the free energy density. We also provide a careful
discussion of the implications of the Mermin-Wagner theorem in the present context and thereby put our low-temperature expansions on safe
grounds.


\maketitle

\section{Introduction}
\label{Intro}

The question of how the spin-wave interaction in a three-dimensional ideal ferromagnet manifests itself in the low-temperature expansion
of the partition function has a very long history. Dyson rigorously answered this question \citep{Dys56}, pointing out errors in some
unsuccessful earlier attempts \citep{Kra36,Ope37,Sch54,Kra55}. After his monumental work, many researchers focused on how to derive
Dyson's result -- which is based on a fairly complicated mathematical formalism -- with alternative methods in a simpler way. Out of the
numerous articles we would like to mention the reference by Zittarz \citep{Zit65}, which solves the problem in a {\it simple} and
{\it elegant} manner, as Dyson put it \citep{Dys96}. More recently, within the effective Lagrangian framework, Dyson's low-temperature
series was rederived \citep{Hof02} and extended to higher orders \citep{Hof11a}. In particular, the general structure of the
low-temperature series of the partition function for a three-dimensional ideal ferromagnet was discussed in the latter reference in full
detail.

Our motivation to write the present article is based on the fact that, apart from some scarce papers distributed over the years, no such
systematic investigation appears to exist in the case of two-dimensional ideal ferromagnets. Above all, to the best of our knowledge, the
effect of the spin-wave interaction on the low-temperature series for the partition function of two-dimensional ferromagnets has never
been addressed so far. The few available papers, all dealing with noninteracting spin waves
\citep{ML69,Col72,YK73,Tak86,Tak87a,Tak87b,Tak90,AA90,SSI94,NT94}, imply that the free energy density -- for a square lattice and in
the absence of a magnetic field -- obeys the following series
\begin{equation}
\label{freeenergyDensitySeries}
z = - {\tilde \eta}_0 T^2 - {\tilde \eta}_1 T^3 + {\cal O}(T^4) \, .
\end{equation}
However, it has never been discussed whether the spin-wave interaction already shows up at order $T^3$ or rather beyond. In other words,
it remains unclear whether the above series referring to the ideal magnon gas is indeed complete up to order $T^3$. Moreover, so far it
has never been discussed in a systematic manner how a weak external magnetic field manifest itself in the above low-temperature series or
how the series looks like for underlying geometries other than a square lattice.

In the present work, using the model-independent and universal method of effective Lagrangians, we systematically evaluate the partition
function of the two-dimensional ideal ferromagnet without restoring to any approximations. We fully take into account lattice anisotropies
which will start manifesting themselves at order $T^3$ in the above series and thereby extend the existing results which strictly apply to
the square lattice by considering also the honeycomb, the triangular and the Kagom\'e lattice. We then show that, up to the order
considered in the above series for the free energy density, we are dealing with noninteracting spin waves -- the interaction sets in only
at order $T^4 \ln T$.

Even in the simple case of noninteracting spin waves, the range of validity of the above low-temperature series derived within the
framework of modified spin-wave theory \citep{Tak86,Tak87a,Tak87b,Tak90} -- which resorts to an ad hoc assumption -- has never been
critically examined. In fact, in Ref.~\citep{PM98} it is stated that {\it to systematically calculate the thermodynamic properties of a
two-dimensional quantum ferromagnet at low temperatures remains an unsolved problem of the spin-wave theory.} To the best of our
knowledge, a rigorous justification of the validity of the results obtained within the framework of modified spin-wave theory is indeed
still lacking. In the present paper not only will we derive the low-temperature properties of two-dimensional ideal ferromagnets in a
systematic manner by using effective Lagrangians, but also we will put our low-temperature series on a firm basis by discussing in detail
the implications of the Mermin-Wagner theorem.

It will also prove to be very instructive to compare the present results for the two-dimensional ferromagnet with those for the
three-dimensional ferromagnet, adopting thereby a unified perspective based on symmetry. In particular, we will discuss the general
structure of the low-temperature series for the free energy density, pointing out how the spin-wave interaction manifests itself in either
case.

The rest of the paper is organized as follows. In Sec.~\ref{EFT} we provide a brief outline of the effective Lagrangian method -- much
more detailed accounts can be found in the pedagogic references \citep{Bur07,Brau10,Leu95,Sch03,Goi04}. The evaluation of the partition
function up to two-loop order in the low-temperature expansion is presented in Sec.~\ref{PartitionFunction} and the thermodynamic
properties of two-dimensional ideal ferromagnets are discussed. The range of validity of the low-temperature series obtained is critically
examined in Sec.~\ref{MerminWagner}. Differences and analogies between two-dimensional and three-dimensional ideal ferromagnets are
discussed in Sec.~\ref{Comparison}. Finally, Sec.~\ref{Summary} contains our conclusions.

We would like to mention that the systematic and model-independent effective Lagrangian method has successfully been applied to other
condensed matter problems. These include antiferromagnets and ferromagnets in two and three spatial dimensions
\citep{HL90,HN91,HN93,Leu94a,Hof99a,Hof99b,RS99a,RS99b,RS00,Hof01,Hof10,Hof11b,Hof11c},
as well as two-dimensional antiferromagnets which are the precursors of high-temperature superconductors
\citep{KMW05,BKMPW06,BKPW06,BHKPW07,BHKMPW07,JKHW09,BHKMPW09,KBWHJW12,VHJW12}.

Also, the correctness of the effective Lagrangian approach was demonstrated explicitly in a recent article on an analytically solvable
microscopic model for a hole-doped ferromagnet in 1+1 dimensions \citep{GHKW10}, by comparing the effective theory predictions with
the microscopic calculation. Furthermore, in a series of high-accuracy numerical investigations of the antiferromagnetic
spin-$\frac{1}{2}$ quantum Heisenberg model on a square lattice using the loop-cluster algorithm \citep{WJ94,GHJNW09,JW11,GHJPSW11}, the
Monte Carlo data were confronted with the analytic predictions of the effective Lagrangian theory and the low-energy constants were
extracted with permille accuracy. All these different tests unambiguously confirm that the effective Lagrangian technique provides a
rigorous and systematic framework to investigate condensed matter systems which exhibit a spontaneously broken continuous symmetry.

\section{Effective Lagrangians at Finite Temperature}
\label{EFT}

The thermodynamic properties of two-dimensional ferromagnets at low temperatures have been investigated before with microscopic methods,
such as modified spin-wave theory \citep{Tak86} or Schwinger Boson mean field theory \citep{AA90}. The corresponding low-temperature
series for the free energy density amounts to a power expansion in the parameter $T/J$, where $J > 0$ is the exchange integral of the
ferromagnetic Heisenberg model
\begin{equation}
{\cal H} = - J \sum_{n.n.} {\vec S}_m \cdot {\vec S}_n - \mu \sum_n {\vec S}_n \cdot {\vec H} \, , \qquad J=const. \, ,
\end{equation}
augmented by the Zeeman term which includes the magnetic field ${\vec H}=(0,0,H)$. The above Hamiltonian, defined on a two-dimensional
lattice with purely isotropic exchange coupling between nearest neighbors, represents what we call {\it ideal} ferromagnet. 

In the present article, we will pursue quite a different approach, based on a rigorous symmetry analysis, which will allow us to derive
the low-temperature properties of two-dimensional ferromagnets. One of the virtues of the effective Lagrangian method is that it is
completely systematic and model-independent. In order not to be repetitive, here we only provide a brief introduction to the effective
Lagrangian method and its extension to finite temperature -- a rather detailed account on finite-temperature effective Lagrangians can be
found in appendix A of Ref.~\citep{Hof11a} and in the various references given therein.

Whenever a global continuous symmetry breaks down spontaneously, Goldstone bosons emerge as the relevant degrees of freedom at low
energies. The effective Lagrangian formulates the dynamics of the system in terms of these Goldstone bosons \citep{Wei79,GL85,Leu94b}. In
the present case of the two-dimensional ferromagnet, the O(3) spin rotation symmetry of the Heisenberg model is spontaneously broken by
the ground state of the ferromagnet which is invariant only under the group O(2). According to the nonrelativistic Goldstone theorem
\citep{Lan66,GHK68,CN76}, two real magnon fields -- or one physical magnon particle -- then occur in the low-energy spectrum of the
ferromagnet.

Having identified the basic degrees of freedom of the effective theory, one systematically constructs the terms appearing in the effective
Lagrangian, order by order in a derivative expansion. The idea is rather simple: One writes down in a systematic way all terms which are
invariant under the symmetries that have been identified in the underlying theory. In the present case of the Heisenberg model these
symmetries include the spontaneously broken spin rotation symmetry O(3), parity, time reversal, as well as the discrete symmetries of the
square, honeycomb, triangular or Kagom\'e lattice.

The various pieces in the effective Lagrangian can be organized according to the number of space and time derivatives which act on the
Goldstone boson fields. This is what is meant by systematic: At low energies, terms in the effective Lagrangian which contain only a few
derivatives are the dominant ones, while terms with more derivatives are suppressed. The leading-order effective Lagrangian for the ideal
ferromagnet is of order $p^2$ and takes the form \citep{Leu94a}
\begin{equation}
\label{leadingLagrangian}
{\cal L}^2_{eff} = \Sigma \frac{\epsilon_{ab} {\partial}_0 U^a U^b}{1+ U^3} + \Sigma \mu H U^3
- \frac{1}{2} F^2 {\partial}_r U^i {\partial}_r U^i \, .
\end{equation}
The two real components of the magnon field, $U^a (a=1,2)$ are the first two components of the three-dimensional unit vector
$U^i = (U^a, U^3)$. While the derivative structure of the above terms is unambiguously determined by the symmetries of the underlying
theory, the two a priori unknown low-energy constants -- the spontaneous magnetization at zero temperature $\Sigma$, and the constant
$F$ -- have to be determined by experiment, numerical simulation or comparison with the microscopic theory.

The above Lagrangian leads to a quadratic dispersion relation
\begin{equation}
\omega({\vec k}) = \gamma {\vec k}^2 + {\cal O}( { |{\vec k}| }^4) \, , \quad \gamma \equiv \frac{F^2}{\Sigma} \, ,
\end{equation}
obeyed by ferromagnetic magnons. This relation dictates how we have to count time and space derivatives in the systematic effective
expansion: One time derivative (${\partial}_0$) is on the same footing as two space derivatives (${\partial}_r {\partial}_r$), i.e., two
powers of momentum count as only one power of energy or temperature: $k^2 \propto \omega, T$. Note that, at this order, lattice
anisotropies do not yet manifest themselves -- the leading order Lagrangian (\ref{leadingLagrangian}) is invariant under continuous space
rotations, although the underlying square, honeycomb, triangular or Kagom\'e lattices are only invariant under discrete space rotations.

As derived in \citep{Hof02}, the next-to-leading order Lagrangian for the ideal ferromagnet is of order $p^4$ and amounts to
\begin{equation}
\label{Leff4}
{\cal L}^4_{eff} = l_1 {( {\partial}_r U^i {\partial}_r U^i )}^2 + l_2 {( {\partial}_r U^i {\partial}_s U^i )}^2
+ l_3 \Delta U^i \Delta U^i \, ,
\end{equation}
where $\Delta$ denotes the Laplace operator in two spatial dimensions. The next-to-leading order effective Lagrangian involves the three
low-energy coupling constants $l_1, l_2$ and $l_3$.

The evaluation of the partition function in Refs.~\citep{Hof02,Hof11a} was based on the assumption that the O(3) space rotation symmetry,
which is an accidental symmetry of the leading order effective Lagrangian, persists at higher orders in the derivative expansion. Here, we
drop this idealization and hence also consider terms in ${\cal L}^4_{eff}$ which are invariant under the discrete symmetries of the
underlying lattice, but no longer invariant under continuous space rotations. Indeed, in the case of the square lattice, the following
extra term
\begin{equation}
\label{Leff4Aniso}
l_4 \sum_{r=1}^2 {\partial}_r {\partial}_r U^i {\partial}_r {\partial}_r U^i
\end{equation}
has to be included in ${\cal L}^4_{eff}$. Interestingly, in the case of the honeycomb, triangular and Kagom\'e lattice, the discrete 60
degrees rotation symmetries do not permit such a term -- here, it is perfectly legitimate to use the space rotation invariant Lagrangian
(\ref{Leff4}). Note that, for the square lattice, there is an additional contribution with four space derivatives
\begin{equation}
\label{Leff4Irrelevant}
\sum_{r=1}^2 {\partial}_r U^i {\partial}_r U^i {\partial}_r U^k {\partial}_r U^k \, .
\end{equation}
However, as we will show below, terms in ${\cal L}^4_{eff}$ that contain four or even more magnon fields, are irrelevant for the evaluation
of the partition function presented in this work.

In finite-temperature field theory the partition function is represented as a Euclidean functional integral
\begin{equation}
\label{TempExp}
\mbox{Tr} \, [\exp(- {\cal H}/T)] = \int [dU] \, \exp \Big(- {\int}_{\! \! \! {\cal T}} \! \! d^4x \, {\cal L}_{eff}\Big) \, .
\end{equation}
The integration extends over all magnon field configurations which are periodic in the Euclidean time direction
$U({\vec x}, x_4 + \beta) = U({\vec x}, x_4)$, with $\beta \equiv 1/T$. The quantity ${\cal L}_{eff}$ on the right hand side is the
Euclidean form of the effective Lagrangian, which consists of a string of terms
\begin{equation}
{\cal L}_{eff} = {\cal L}^2_{eff} + {\cal L}^4_{eff} + {\cal O}(p^6) \, ,
\end{equation}
involving an increasing number of space and time derivatives.

\begin{figure}
\includegraphics[width=13.5cm]{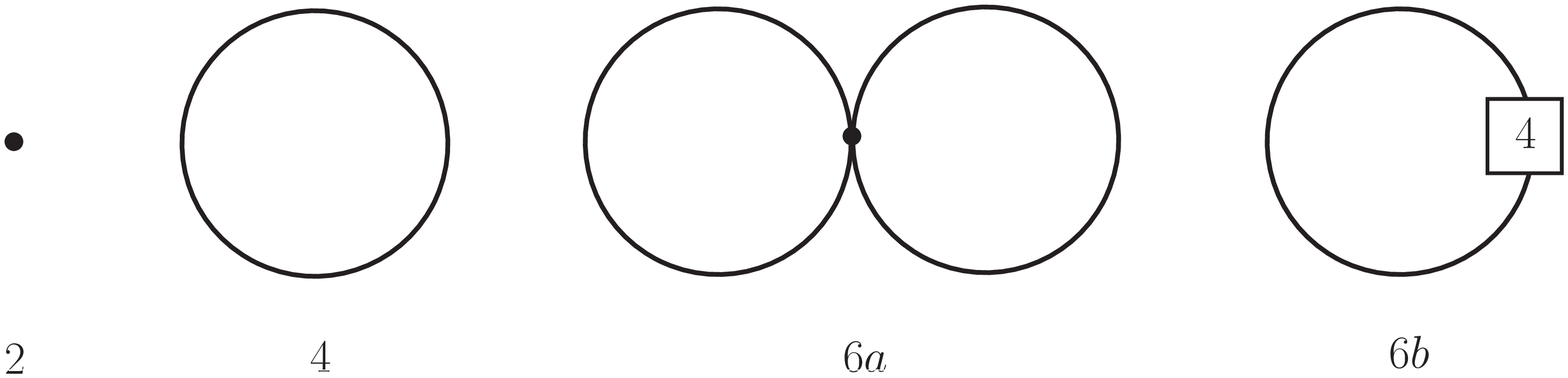}

\vspace{4mm}

\includegraphics[width=13.0cm]{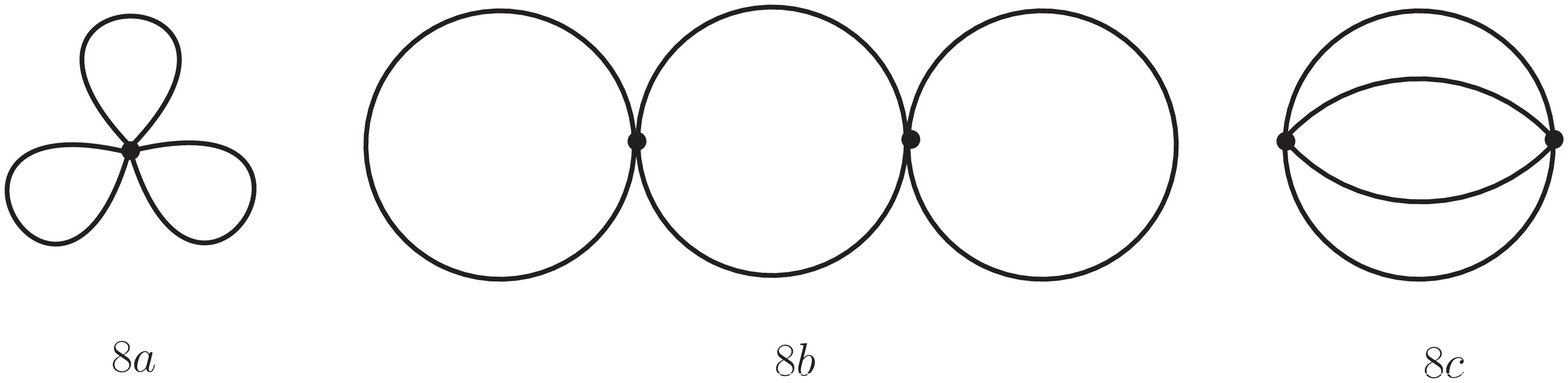}

\vspace{4mm}

\includegraphics[width=10.5cm]{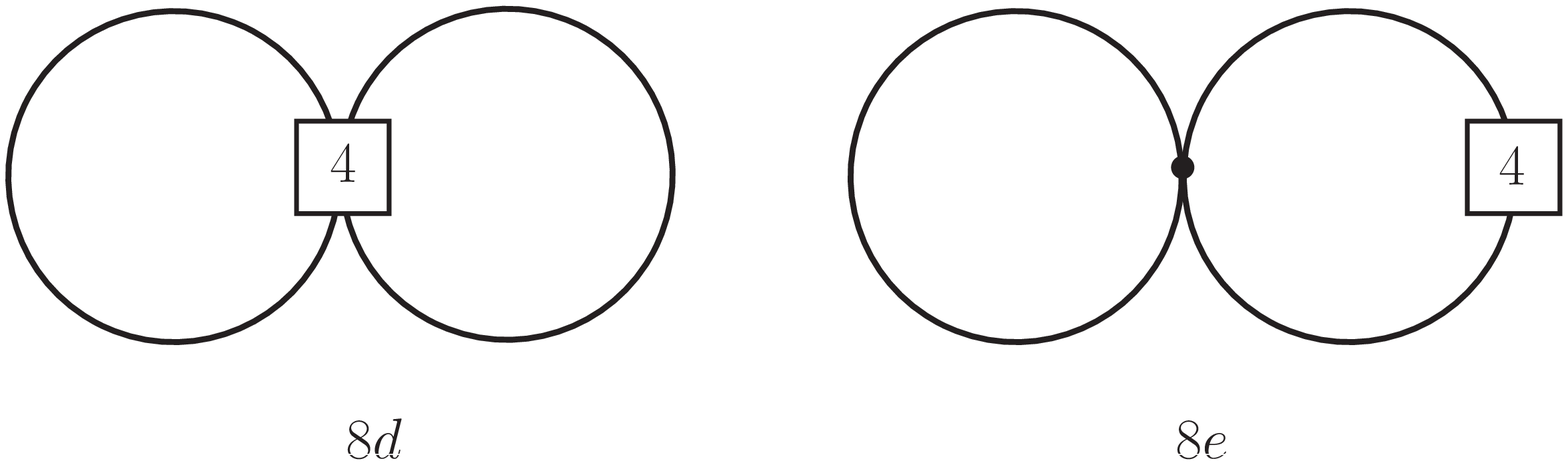}

\vspace{4mm}

\includegraphics[width=7.0cm]{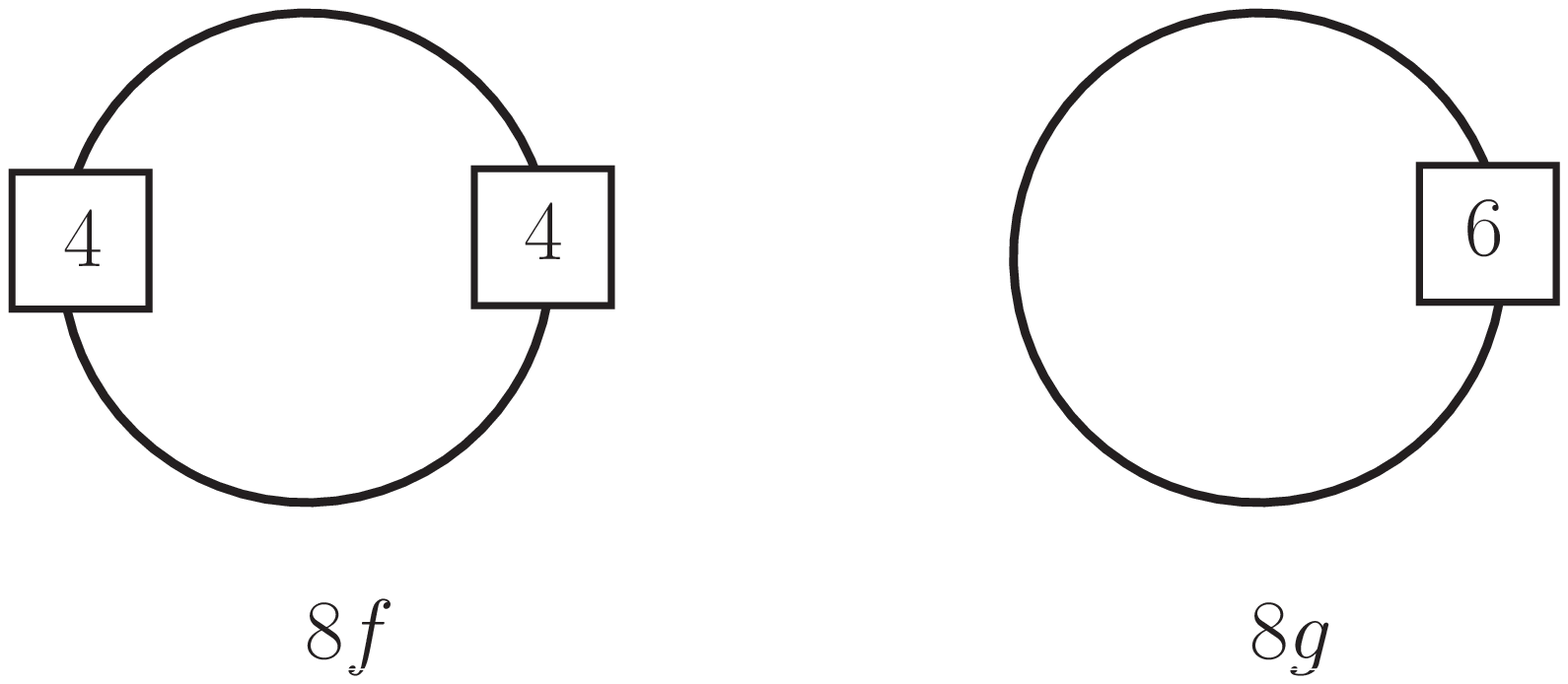}

\caption{Feynman graphs related to the low-temperature expansion of the partition function for a two-dimensional ferromagnet up to order
$p^8$. The numbers attached to the vertices refer to the piece of the effective Lagrangian they come from. Vertices associated with the
leading term ${\cal L}^2_{eff}$ are denoted by a dot. Note that ferromagnetic loops are suppressed by two powers of momentum in $d_s$=2.}
\label{figure1}
\end{figure}

The virtue of the representation (\ref{TempExp}) lies in the fact that it can be evaluated perturbatively. To a given order in the
low-temperature expansion only a finite number of Feynman graphs and only a finite number of effective coupling constants contribute. The
low-temperature expansion of the partition function is obtained by considering the fluctuations of the spontaneous magnetization vector
field ${\vec U} = (U^1, U^2, U^3 )$ around the ground state ${\vec U_0} = (0, 0, 1)$, i.e., by expanding $U^3$ in powers of the spin-wave
fluctuations $U^a$,
\begin{equation}
U^3 = \sqrt{1-U^aU^a} = 1 - \mbox{$\frac{1}{2}$} U^a U^a - \mbox{$\frac{1}{8}$} U^a U^a U^b U^b - \dots \, . 
\end{equation}
Inserting this expansion into formula (\ref{TempExp}) one then generates the Feynman diagrams illustrated in Fig.~\ref{figure1}. The
leading contribution in the exponential on the right hand side of Eq.~(\ref{TempExp}) is of order $p^2$ and originates from
${\cal L}^2_{eff}$. It contains a term quadratic in the spin-wave field $U^a$ -- with the appropriate derivatives and the magnetic field
displayed in Eq.(\ref{leadingLagrangian}) -- and describes free magnons. The corresponding diagram for the partition function is the
one-loop diagram 4 of Fig.~\ref{figure1}.

A crucial point underlying the perturbative evaluation of the partition function of any system concerns the suppression of loop diagrams
in the effective field theory framework. In three spatial dimensions -- and in the case of the ferromagnet -- each loop in a Feynman
diagram is suppressed by three powers of momentum. In two dimensions, on the other hand, ferromagnetic loops are only suppressed by
{\bf two} powers of momentum: The one-loop diagram 4 is of order $p^4$, as it involves ${\cal L}^2_{eff}$ ($p^2$) and one loop ($p^2$).

The reason why the loop suppression depends on the spatial dimension $d_s$ of the system can easily be appreciated as follows: Each loop
involves an integral of the type
\begin{equation}
\int d \omega \, d^{d_s} k \ \frac{1}{\omega - \gamma {\vec k}^2} \ \propto \ p^{d_s} \, ,
\end{equation}
related to ferromagnetic magnons circling in the loop. On dimensional grounds the integral is proportional to $d_s$ powers of momentum,
i.e., each loop in a Feynman diagram referring to the two-dimensional ferromagnet is suppressed by $p^2$.

The remainder of the effective Lagrangian in the path integral formula for the partition function (\ref{TempExp}), i.e.
${\cal L}^4_{eff} + {\cal L}^6_{eff} + \dots$, is treated as a perturbation. The Gaussian integrals are evaluated in the standard manner
(see Ref.~\citep{Kap98}, in particular chapter 3) and one arrives at a set of Feynman rules which differ from the zero-temperature rules
of the effective Lagrangian method only in one respect: the periodicity condition imposed on the magnon fields modifies the propagator. At
finite temperature, the propagator is given by
\begin{equation}
\label{ThermalPropagator}
G(x) = \sum_{n \,= \, - \infty}^{\infty} \Delta({\vec x}, x_4 + n \beta) \, ,
\end{equation}
where $\Delta(x)$ is the Euclidean propagator at zero temperature,
\begin{equation}
\label{Propagator}
\Delta (x) = \int \! \, \frac{d k_4 d^2\!k}{(2\pi)^3} \frac{e^{i{\vec k}{\vec x} - i k_4 x_4}}{\gamma {\vec k}^2 - i k_4 + \mu H}
= \Theta(x_4) \int \! \, \frac{d^2\!k}{(2\pi)^2} \, e^{i{\vec k}{\vec x} - \gamma {\vec k}^2 x_4 -\mu H x_4} \, . 
\end{equation}
An explicit representation for the thermal propagator, dimensionally regularized in the spatial dimension
$d_s$, is
\begin{equation}
\label{ThermProp}
G(x) = \frac{1}{(2{\pi})^{d_s}} \, \Big(\frac{{\pi}}{\gamma}\Big)^{\frac{d_s}{2}} \sum^{\infty}_{n \, = \, - \infty} \frac{1}{x_n^{\frac{d_s}{2}}}
\, \exp^{- \frac{{\vec x}^2}{4 \gamma x_n} - \mu H x_n} \, \Theta (x_n) \, ,
\end{equation}
with
\begin{equation}
x_n \, \equiv \, x_4 + n \beta \, .
\end{equation}

We restrict ourselves to the infinite volume limit and evaluate the free energy density $z$, defined by
\begin{equation}
\label{freeEnergyDensity}
z = - \, T \, \lim_{L\to\infty} L^{-2} \, \ln \, [\mbox{Tr} \exp(-{\cal H}/T)] \, .
\end{equation}

Note again that in the case of a quadratic dispersion relation -- and in two space dimensions -- each loop in a Feynman diagram is
suppressed by {\bf two} powers of momentum. This suppression rule lies at the heart of the organization of the Feynman graphs of the
partition function for the two-dimensional ferromagnet depicted in Fig.~\ref{figure1}. Now we also understand why terms in
${\cal L}^4_{eff}$ that contain four or even more magnon fields are irrelevant for the explicit evaluation of the partition function
presented in this work which goes up to order $p^6$: The two-loop diagram 8d with an insertion from ${\cal L}^4_{eff}$, containing four
magnon fields, is of order $p^8$, as it involves ${\cal L}^4_{eff}$ ($p^4$) and two loops ($p^4$).

In the next section we will evaluate the partition function of the two-dimensional ideal ferromagnet in full generality up to order $p^6$.
The evaluation of the partition function at order $p^8$ is much more involved. In particular, the renormalization and numerical evaluation
of the three-loop graph 8c turns out to be rather elaborate -- a detailed account of this calculation will be presented elsewhere
\citep{Hof12}. Here, we rather focus on the general structure of the low-temperature expansion and answer the question which contributions
originate from noninteracting spin waves and which ones are due to the spin-wave interaction -- this question has never been addressed so
far.

\section{Thermodynamics of Two-Dimensional Ideal Ferromagnets}
\label{PartitionFunction}

We now consider those Feynman graphs depicted in Fig.~\ref{figure1} that contribute to the partition function up to order $p^6$ or,
equivalently, up to order $T^3$. Again, additional information on finite-temperature effective Lagrangians and the evaluation of the
corresponding Feynman diagrams -- going beyond the outline given in the previous section -- can be found in Ref.~\citep{Hof11a} (see
section III and appendix A).

At leading order (order $p^2$), we have the tree graph 2 involving ${\cal L}^2_{eff}$ which merely leads to a temperature-independent
contribution to the free energy density,
\begin{equation}
z_{2} = - \Sigma \mu H \, .
\end{equation}

The leading temperature-dependent contribution is of order $p^4$ and stems from the one-loop graph 4. It is associated with a
$(d_s+1)$-dimensional nonrelativistic free Bose gas and amounts to
\begin{equation}
\label{z(4)T}
z^T_{4} = - \frac{1}{4 \pi {\gamma}} \, T^2 \sum^{\infty}_{n=1} \frac{e^{- \mu H n \beta}}{n^2} \, .
\end{equation}

At order $p^6$ the first two-loop graph shows up. This contribution, related to graph 6a, is proportional to single space derivatives of
the propagator at the origin,
\begin{equation}
\label{z(6a)}
z_{6a} \ \propto \, \Big[ {\partial}_r G(x) \Big]_{x=0} \, \Big[{\partial}_r G(x) \Big]_{x=0} = 0 \, ,
\end{equation}
and thus vanishes because the thermal propagator is invariant under parity, much like the Heisenberg Hamiltonian. Remember that the
effective Lagrangian - and therefore the thermal propagator -- inherits all the symmetries of the underlying Heisenberg model.

At the same order $p^6$, the next-to-leading order Lagrangian ${\cal L}^4_{eff}$ comes into play. The one-loop graph 6b, which involves a
two-magnon vertex, corresponds to
\begin{equation}
\label{z(6b)}
z_{6b} = - \frac{2 \, l_3}{{\Sigma}} \, \Big[ {\Delta}^2 G(x) \Big]_{x=0}
- \frac{2 \, l_4}{{\Sigma}} \, \Big[ \sum_{r=1}^2 {\partial}^4_r G(x) \Big]_{x=0} \, ,
\end{equation}
and yields the temperature-dependent contribution
\begin{equation}
\label{z(6b)T}
z^T_{6b} = - \frac{4 l_3 + 3 l_4}{4 \pi \Sigma {\gamma}^3} \, T^3 \sum^{\infty}_{n=1} \frac{e^{- \mu H n \beta}}{n^3} \, .
\end{equation}

Collecting terms, the free energy density of the two-dimensional ideal ferromagnet up to order $p^6 \propto T^3$ becomes
\begin{equation}
\label{FreeCollect}
z = - \Sigma \mu H - \frac{1}{4 \pi \gamma} \, T^2 \, \sum^{\infty}_{n=1} \frac{e^{- \mu H n \beta}}{n^2}
- \frac{4 l_3 + 3 l_4}{4 \pi \Sigma {\gamma}^3} \, T^3 \, \sum^{\infty}_{n=1} \frac{e^{- \mu H n \beta}}{n^3} + {\cal O}(p^8) \, .
\end{equation}
The contributions of order $T^2$ and $T^3$ arise from one-loop graphs and are related to the free energy density of noninteracting spin
waves. The former contribution is exclusively determined by the leading-order effective constants $\Sigma$ and $F$
($\gamma = F^2/\Sigma$), i.e., it is the same for any of the four types of lattices -- square, honeycomb, triangular and Kagom\'e --
considered here. The term of order $T^3$, on the other hand, is not universal since it involves the next-to-leading order effective
constants $l_3$ and $l_4$ from ${\cal L}^4_{eff}$. In the case of the honeycomb, triangular and Kagom\'e lattice, the coefficient of
order $T^3$ exclusively contains the contribution from $l_3$ -- the term (\ref{Leff4Aniso}) in the effective Lagrangian involving $l_4$,
which accounts for the lattice anisotropies, is excluded due to the discrete 60 degrees rotation symmetries of these lattices. Remarkably,
the spin-wave interaction does not yet manifest itself at this order in the low-temperature expansion of the free energy density. The only
potential candidate, the two-loop diagram 6a of order $T^3$, turns out to be zero due to parity.

The ratio $\mu H \beta = \mu H / T$ in the above series can take any value, as long as the temperature and the magnetic field themselves
are small compared to the intrinsic scale of the underlying theory, which in the present case of the two-dimensional ferromagnet is given
by the exchange integral $J$ of the Heisenberg model. In the following we will be interested in the limit $T \gg \mu H$ which we implement
by holding $T$ fixed and sending the magnetic field to zero. Since we keep the fixed temperature small compared to the scale $J$, we never
leave the domain of validity of the low-temperature expansion.

In order to discuss the effect of a weak magnetic field we thus expand the result (\ref{FreeCollect}) in the dimensionless parameter
\begin{equation}
\sigma = \mu H \beta = \frac{\mu H}{T} \, .
\end{equation}
Retaining all terms up to quadratic in $\sigma$, we obtain
\begin{eqnarray}
\label{FreeCollectSmallH}
z & = & - \Sigma \mu H - \frac{1}{4 \pi \gamma} \, T^2 \, \Big\{ \zeta(2) + \sigma \ln \sigma - \sigma
- \frac{{\sigma}^2}{4} + {\cal O}({\sigma}^3) \Big\} \nonumber \\
& & - \frac{4 l_3 + 3 l_4}{4 \pi \Sigma {\gamma}^{3}} \, T^3 \, \Big\{ \zeta(3)  - \zeta(2) \, \sigma - \frac{1}{2} {\sigma}^2 \ln \sigma
+ \frac{3 {\sigma}^2}{4} + {\cal O}({\sigma}^3) \Big\} + {\cal O}(p^8) \, .
\end{eqnarray}
In the absence of an external magnetic field, the sums in the series (\ref{FreeCollect}) become temperature independent and reduce to
Riemann zeta functions,
\begin{equation}
\label{FreeCollectStructure}
z = - \frac{1}{4 \pi \gamma} \, \zeta(2) \, T^2 - \frac{4 l_3 + 3 l_4}{4 \pi \Sigma {\gamma}^{3}} \, \zeta(3) \, T^3
+ {\cal O}(p^8) \, .
\end{equation}
Since we are dealing with a two-dimensional system, we have to be careful by taking the limit $H \! \to \! 0$ due to the Mermin-Wagner
theorem. A thorough discussion, confirming the validity of the above series, will be given in Sec.~\ref{MerminWagner}.

To corroborate the structure of the low-temperature series, let us consider an independent derivation, based on the evaluation of the
two-point function and the subsequent extraction of the dispersion relation. The free energy density of a two-dimensional gas of
noninteracting bosons is then obtained from the dispersion relation through
\begin{equation}
\label{FreeFunction2}
z = z_0 \, + \, \frac{T}{(2 \pi)^2} \, \int \!\! d^2 \! k \, \ln \Big[ 1 - e^{- \omega({\vec k}) / T} \Big] \, ,
\end{equation}
where $z_0$ is the free energy density of the vacuum. The leading term in the dispersion relation,
\begin{equation}
\omega(\vec k) = \gamma{\vec k}^2 + \mu H + {\cal O}(|{\vec k}|^4) \, , \qquad \quad \gamma \equiv \frac{F^2}{\Sigma} \, ,
\end{equation}
yields the dominant one-loop contribution $z^T_4 \propto T^2$ in the free energy density of the two-dimensional ferromagnet.

\begin{figure}
\includegraphics[width=12cm]{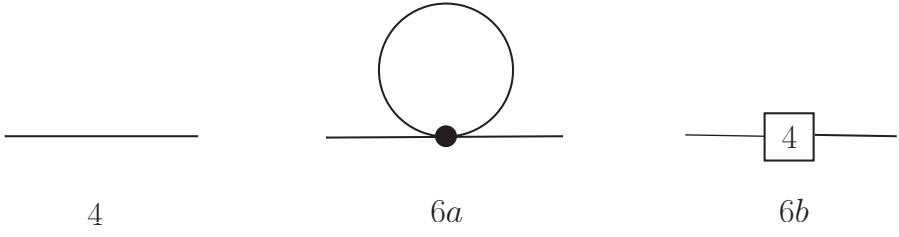}
\caption{Feynman graphs occurring in the low-energy expansion of the two-point function for a two-dimensional ferromagnet up to order
$p^6$.}
\label{figure2}
\end{figure}

Subleading terms in the dispersion relation are obtained by evaluating the two-point function to higher orders. The relevant graphs are
shown in Fig.~\ref{figure2}. Depicted are all contributions up to order $p^6$. Instead of listing individual results for the two-point
function, we give the final expression for the dispersion relation originating from these graphs:
\begin{equation}
\label{DispRel}
\omega(\vec k) = \frac{1}{\Sigma} \, \Big\{ \Sigma \mu H + F^2 {\vec k}^2 - (2 l_3 + \mbox{$\frac{3}{2}$} l_4) {\vec k}^4
+ {\cal O}({\vec k}^6) \Big\} \, .
\end{equation}
Note that the one-loop graph 6a does not contribute to the dispersion relation. There is thus only one additional diagram contributing
at the order we are considering: Graph 6b which leads to a higher-order term involving the effective constants $l_3$ and $l_4$. Again, in
the case of the honeycomb, triangular, and Kagom\'e lattices, the contribution proportional to the low-energy constant $l_4$ is absent.
Inserting the above expression into the free Bose gas formula (\ref{FreeFunction2}) one readily confirms the low-temperature series for
the free energy density.

Let us also consider the low-temperature series for the energy density $u$, for the entropy density $s$, and for the heat capacity $c_V$
of the two-dimensional ideal ferromagnet. They are readily worked out from the thermodynamic relations
\begin{equation}
\label{Thermodynamics}
s = \frac{{\partial}P}{{\partial}T} \, , \qquad u = Ts - P \, , \qquad 
c_V = \frac{{\partial}u}{{\partial}T} = T \, \frac{{\partial}s}{{\partial}T} \, .
\end{equation}
Because the system is homogeneous, the pressure can be obtained from the temperature-dependent part of the free energy density,
\begin{equation}
\label{Pz}
P = z_0 - z \, ,
\end{equation}
such that the other thermodynamic quantities amount to
\begin{eqnarray}
u & = & \frac{1}{4 \pi \gamma} \, T^2 \, \Bigg\{ \sigma \sum^{\infty}_{n=1} \frac{e^{- \sigma n}}{n}
+ \sum^{\infty}_{n=1} \frac{e^{- \sigma n}}{n^2} \Bigg\} \nonumber \\
& & + \frac{4 l_3 + 3 l_4}{4 \pi \Sigma {\gamma}^3} \, T^3 \, \Bigg\{ \sigma \sum^{\infty}_{n=1} \frac{e^{- \sigma n}}{n^2}
+ 2 \sum^{\infty}_{n=1} \frac{e^{- \sigma n}}{n^3} \Bigg\}
+ {\cal O}(p^8) \, , \nonumber \\
s & = & \frac{1}{4 \pi \gamma} \, T \, \Bigg\{ \sigma \sum^{\infty}_{n=1} \frac{e^{- \sigma n}}{n}
+ 2 \sum^{\infty}_{n=1} \frac{e^{- \sigma n}}{n^2} \Bigg\} \nonumber \\
& & + \frac{4 l_3 + 3 l_4}{4 \pi \Sigma {\gamma}^3} \, T^2 \, \Bigg\{ \sigma \sum^{\infty}_{n=1} \frac{e^{- \sigma n}}{n^2}
+ 3 \sum^{\infty}_{n=1} \frac{e^{- \sigma n}}{n^3} \Bigg\}
+ {\cal O}(p^6) \, , \nonumber \\
c_V & = & \frac{1}{4 \pi \gamma} \, T \, \Bigg\{ {\sigma}^2 \sum^{\infty}_{n=1} e^{- \sigma n}
+ 2 \sigma \sum^{\infty}_{n=1} \frac{e^{- \sigma n}}{n}
+ 2 \sum^{\infty}_{n=1} \frac{e^{- \sigma n}}{n^2} \Bigg\} \nonumber \\
& & + \frac{4 l_3 + 3 l_4}{4 \pi \Sigma {\gamma}^3} \, T^2 \, \Bigg\{ {\sigma}^2 \sum^{\infty}_{n=1} \frac{e^{- \sigma n}}{n}
+ 4 \sigma \sum^{\infty}_{n=1} \frac{e^{- \sigma n}}{n^2}
+ 6 \sum^{\infty}_{n=1} \frac{e^{- \sigma n}}{n^3} \Bigg\}
+ {\cal O}(p^6) \, .
\end{eqnarray}
For a weak magnetic field $H$, the series may be expanded in the parameter $\sigma = \mu H /T$,
\begin{eqnarray}
u & = & \frac{1}{4 \pi \gamma} \, T^2 \, \Big\{ \zeta(2) - \sigma + \frac{{\sigma}^2}{4} + {\cal O}({\sigma}^3) \Big\} \nonumber \\
& & + \frac{4 l_3 + 3 l_4}{4 \pi \Sigma {\gamma}^{3}} \, T^3 \, \Big\{ 2 \zeta(3) - \zeta(2) \sigma + \frac{{\sigma}^2}{2}
+ {\cal O}({\sigma}^3) \Big\}
+ {\cal O}(p^8) \, ,
\nonumber \\
s & = & \frac{1}{4 \pi \gamma} \, T \, \Big\{ 2 \zeta(2) + \sigma \ln \sigma - 2 {\sigma} + {\cal O}({\sigma}^3) \Big\} \nonumber \\
& & + \frac{4 l_3 + 3 l_4}{4 \pi \Sigma {\gamma}^{3}} \, T^2 \, \Big\{ 3 \zeta(3) - 2 \zeta(2) \sigma - \frac{{\sigma}^2}{2}
\ln \sigma + \frac{5{\sigma}^2}{4} + {\cal O}({\sigma}^4) \Big\}
+ {\cal O}(p^6) \, ,
\nonumber \\
c_V & = & \frac{1}{4 \pi \gamma} \, T \, \Big\{ 2 \zeta(2) - {\sigma} + {\cal O}({\sigma}^3) \Big\} \nonumber \\
& & + \frac{4 l_3 + 3 l_4}{4 \pi \Sigma {\gamma}^{3}} \, T^2 \, \Big\{ 6 \zeta(3) - 2 \zeta(2) {\sigma} + \frac{{\sigma}^2}{2}
+ {\cal O}({\sigma}^4) \Big\}
+ {\cal O}(p^6) \, ,
\end{eqnarray}
where we have retained terms up to quadratic in the magnetic field. Formally, as it was the case for the free energy density, the limit
$H \! \to \! 0$ poses no problems. Note again that all contributions in the above series for $u$, $s$ and $c_V$ originate from one-loop
graphs -- the spin-wave interaction does not yet manifest itself at this order of the low-temperature expansion.

We may now compare our results, derived within the effective field theory framework, with the literature. The few explicit results
available all refer to the limit $H \! \to \! 0$ and to the square lattice \citep{ML69,Tak86,Tak87a,Tak87b,Tak90,AA90}. For the free
energy density of the two-dimensional ideal ferromagnet these authors obtain
\begin{equation}
z = - \frac{1}{4 \pi J S a^2} \, \zeta(2) \, T^2 - \frac{1}{32 \pi J^2 S^2 a^2} \, \zeta(3) \, T^3 + {\cal O}(T^4) \, .
\end{equation}
The first term coincides with our series (\ref{FreeCollectStructure}) provided that we express the effective low-energy constant $\gamma$
in terms of microscopic constants as
\begin{equation}
\label{gammamicro}
\gamma = J S a^2 \, .
\end{equation}
By matching the coefficients of the second term, the combination of the low-energy constants $l_3$ and $l_4$ is identified as
\begin{equation}
l_3 + \mbox{$\frac{3}{4}$} l_4 = \frac{J S^2 a^2}{32} \, .
\end{equation}
The microscopic calculation is thus consistent with the effective calculation for the square lattice in the limit $H \! \to \! 0$.

However, the author of Ref.~\citep{Tak86}, which is considered as a standard reference on the low-temperature properties of
two-dimensional ferromagnets, was rather cautious about the correctness or validity of his result: Regarding the low-temperature series
for the free energy density he comments that it is {\it possible} that his series {\it gives the correct low-temperature expansion.}

The reason for his caution may be readily identified. While spin-wave theory works well for three-dimensional systems, in two or one space
dimensions the spin-wave expansion is plagued with divergences. In order to cope with low-dimensional systems, many approximations were
invented. One very popular method is modified spin-wave theory, advocated first for two-dimensional Heisenberg ferromagnets \citep{Tak86}
and then transferred to two-dimensional antiferromagnets \citep{Tak89,HT89,TLH89}. The essential idea is to impose an ad hoc condition on
the chemical potential. However, to the best of our knowledge, the justification of such an ad hoc condition was never rigorously
examined, neither for the ferromagnet nor for the antiferromagnet.

Within our effective field theory framework, we will put the above low-temperature series for a two-dimensional system on a firm basis --
on the same footing as the low-temperature series for ferro- and antiferromagnets in three space dimensions. Indeed, as we will show in
the next section, the Mermin-Wagner theorem is perfectly consistent with the low-temperature series derived in the present work. We also
like to emphasize that the effective field theory approach we have pursued does not resort to any approximations or ad hoc conditions as
e.g. in the case of modified spin-wave theory. Moreover, our series go beyond the results of the literature, as they explicitly include a
weak external magnetic field and are valid not only for the square lattice, but also for the honeycomb, the triangular and the Kagom\'e
lattice with a spontaneously broken spin symmetry O(3) $\to$ O(2).

\section{Range of Validity of the Low-Temperature Series}
\label{MerminWagner}

Whereas in three space dimensions the limit ${H \! \to \! 0}$ can readily be taken, one has to be careful in two (or one) dimensions due to
the Mermin-Wagner theorem \citep{MW68}, which states that no spontaneous symmetry breaking at any finite temperature in the O(3)-invariant
two-dimensional Heisenberg model can occur. Accordingly, no massless magnons in the low-energy spectrum at any finite temperature will be
present. In the context of ferromagnetic magnons this means that the low-energy spectrum exhibits a nonperturbatively generated energy gap
and that the correlation length of the magnons no longer is infinite. Still, the correlation length is exponentially large, the argument
of the exponential being proportional to the inverse temperature \citep{KC89},
\begin{equation}
\label{npcorrelation}
\xi_{np} = C_{\xi} a S^{-\frac{1}{2}} \, \sqrt{\frac{T}{J S^2}} \, \exp \! \Big[\frac{2 \pi J S^2}{T} \Big] \, .
\end{equation}
Here $a$ is the spacing between two neighboring sites on the square lattice and the quantity $ C_{\xi} \approx 0.05$ is a dimensionless
constant.

Strictly speaking, it is therefore not legitimate to switch off the external magnetic field $H$ completely, because our effective
calculation does not take into account the nonperturbative effect. However, the corrections due to the nonperturbatively generated
energy gap are so tiny, that they cannot manifest themselves in the power series derived in this work. In what follows, we will estimate
the order of magnitude of these corrections and thus verify this claim. While the above explicit expression for the correlation length
refers to the square lattice, note that the conclusions to be presented in this section also apply to the honeycomb, the triangular and
the Kagom\'e lattice. 

Our low-temperature series are valid as long as the correlation length $\xi$ of the Goldstone bosons is much smaller than the
nonperturbatively generated correlation length $\xi_{np}$. In order to define the correlation length $\xi$ for ferromagnetic magnons in a
natural way, let us consider the dispersion relation. In the presence of a magnetic field it takes the form
\begin{equation}
\omega({\vec k}) = \gamma {\vec k}^2 + \mu H + {\cal O}( { |{\vec k}| }^4) \, , \quad \gamma \equiv \frac{F^2}{\Sigma} \, ,
\end{equation}
and we may define the correlation length as
\begin{equation}
\xi = \sqrt{\frac{\gamma}{\mu H}} = \frac{F}{\sqrt{\Sigma\mu H}} \, .
\end{equation}
This quantity has dimension of length and tends to infinity for zero magnetic field, as it should. It is the analog of the corresponding
relation for antiferromagnetic magnons, which obey a linear (relativistic) dispersion law. In that case the correlation length is given by
the inverse mass $M_{\pi}$ \citep{Hof10},
\begin{equation}
{\xi}_{AF} = \frac{1}{M_{\pi}} = \frac{F_{AF}}{\sqrt{\Sigma_s \mu H_s}} \, ,
\end{equation}
where $\Sigma_s$ and $H_s$ are the staggered magnetization at zero temperature and the staggered field, respectively. 
 
The low-temperature series derived in the previous section are certainly valid if $\xi_{np}$ is -- let us say -- a thousand times larger
than $\xi$, i.e.
\begin{equation}
\label{npestimate}
\frac{1}{1000} = \frac{\xi}{\xi_{np}}
= \frac{S^2}{C_{\xi}} \, \frac{J}{T} \, \sqrt{\frac{T}{\mu H}} \,
\exp \! \Big[- \frac{2 \pi J S^2}{T} \Big] \, .
\end{equation}
Note that we have used eq.~(\ref{gammamicro}) in order to express the effective constant $F$ in terms of the exchange integral $J$ of the
underlying theory by
\begin{equation}
F = \sqrt{\Sigma J S} a \, .
\end{equation}
Now the exchange integral defines a scale in the underlying theory and for the effective expansion to be consistent, the temperature has
to be small with respect to this scale. Assuming that
\begin{equation}
\frac{T}{J} = \frac{1}{100} \, ,
\end{equation}
relation (\ref{npestimate}) then yields the ratio
\begin{equation}
\frac{\mu H}{T} \approx 10^{-125} \qquad (S =\mbox{$\frac{1}{2}$}) \, .
\end{equation}
We thus see that, in principle, we cannot completely switch off the magnetic field -- rather, we start running into trouble as soon as the
ratio $\mu H /T$ is of the order of the above value. However, the error introduced into the low-temperature series considered in this work
is indeed extremely small. Hence we confirm that the corrections due to the nonperturbatively generated energy gap are so tiny that they
cannot manifest themselves in the above low-temperature expansions for the thermodynamic quantities -- the subtleties raised by the
Mermin-Wagner theorem in $d_s$=2 are not relevant for our calculation.

The effective calculation performed in this work is restricted to the regime $\xi \ll \xi_{np}$. This does not mean that the regime
$\xi \gg \xi_{np}$ is beyond the reach of the effective field theory. Rather, one has to resort to a different type of perturbative
expansion. A similar situation arises when one considers finite-size effects. In particular, when the Goldstone boson mass is small
compared to the inverse size of the box, a different effective expansion scheme, the so-called $\epsilon$-expansion, applies. Indeed,
various problems within this framework have been investigated in detail \citep{GL87,Leu87,GL88,GL91a,GL91b,DNJN91}.

We close this section with a conceptual remark. Our effective analysis refers to the two-dimensional {\it ideal} ferromagnet, i.e., to a
two-dimensional system which is governed by the isotropic exchange interaction and the interaction with a weak external magnetic field.
This represents the system which was analyzed before within a microscopic framework by Takahashi and other authors
\citep{ML69,Col72,YK73,Tak86,Tak87a,Tak87b,Tak90,AA90,SSI94,NT94}. For this idealized situation we have rigorously shown that taking the
limit $H \! \to \! 0$ in the low-temperature series derived in this work is consistent with the Mermin-Wagner theorem.

In a more realistic approach to ferromagnetic films, however, one has to also consider the magnetic anisotropy and dipolar interactions.
Although they are much weaker than the exchange interaction, these effects may play a decisive role (see e.g.~\citep{Bru92}). In
particular, taking into account these effects, the Mermin-Wagner theorem is evaded since two of its basic assumptions are no longer
fulfilled: the Hamiltonian is no longer isotropic and the dipolar interaction is no longer short-ranged.

So the question regarding the implications of the Mermin-Wagner theorem on the low-temperature properties of a two-dimensional ideal
ferromagnet is rather academic. Still, unlike the various authors before
\citep{ML69,Col72,YK73,Tak86,Tak87a,Tak87b,Tak90,AA90,SSI94,NT94}, here we have put the low-temperature series on a firm basis in this
idealized framework.

\section{Ideal Ferromagnets in Two and Three Spatial Dimensions -- A Comparison}
\label{Comparison}

It is very instructive to compare the thermodynamic properties of two-dimensional ferromagnets with those of three-dimensional ideal
ferromagnets within the effective field theory perspective, adopting thereby a unified point of view based on the symmetries of the
system.

\begin{figure}
\includegraphics[width=13.5cm]{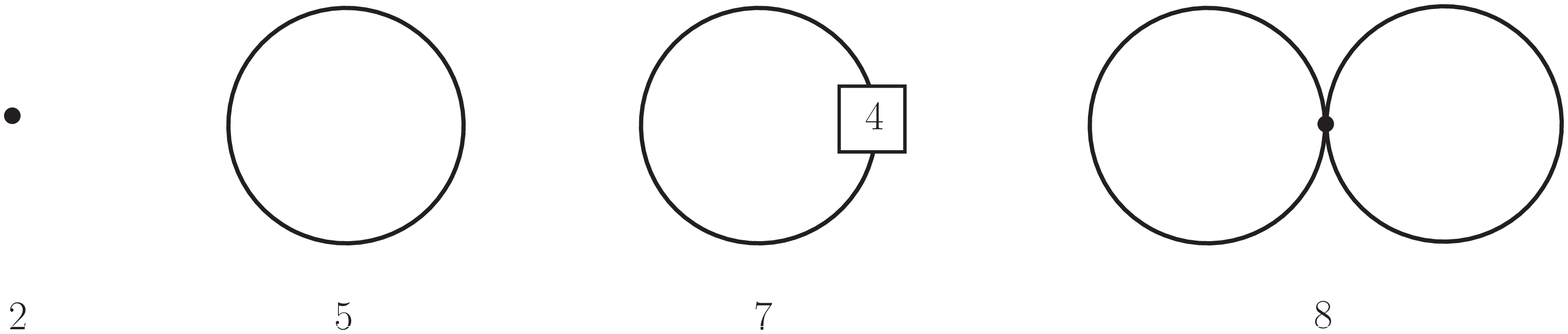}

\vspace{4mm}

\includegraphics[width=16.0cm]{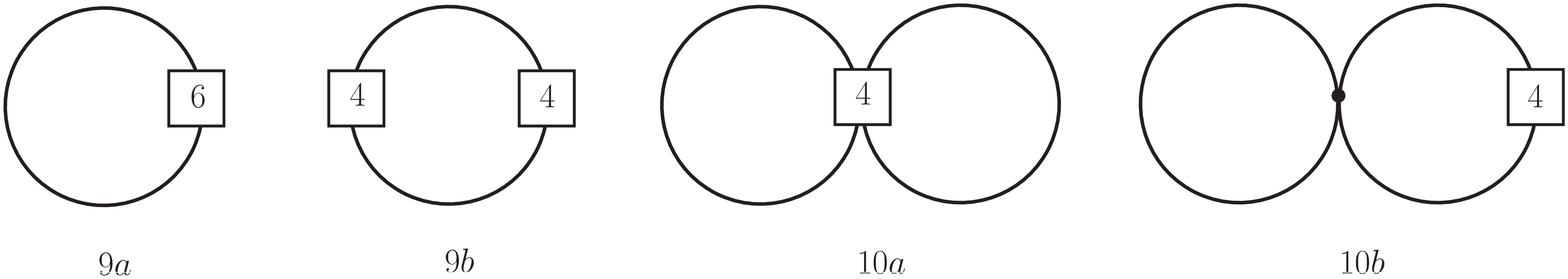}

\vspace{4mm}

\includegraphics[width=11.5cm]{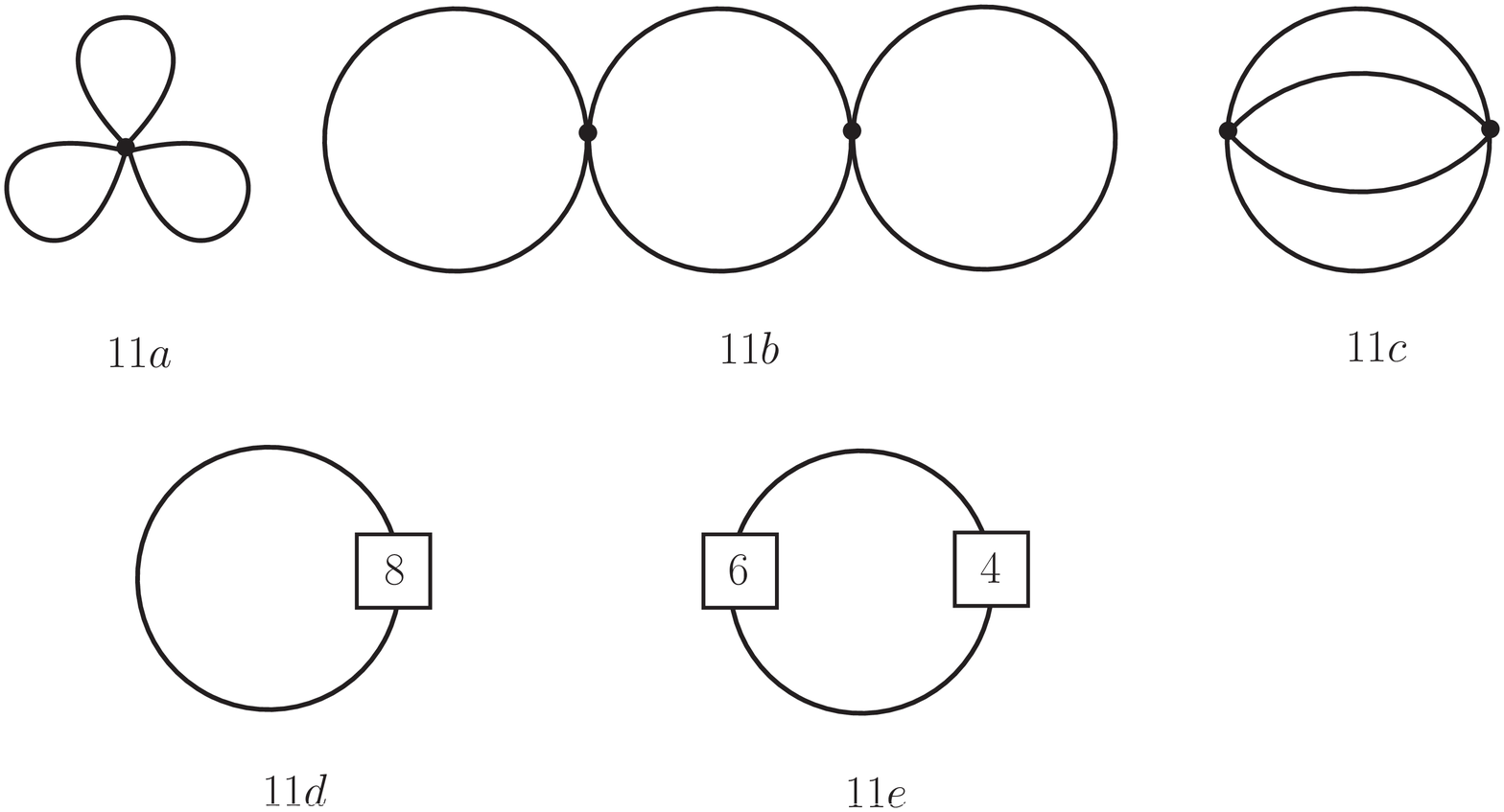}

\caption{Feynman graphs related to the low-temperature expansion of the partition function for a {\bf three-dimensional ferromagnet} up to
order $p^{11}$. The numbers attached to the vertices refer to the piece of the effective Lagrangian they come from. Vertices associated
with the leading term ${\cal L}^2_{eff}$ are denoted by a dot. Note that ferromagnetic loops are suppressed by three powers of momentum in
$d_s$=3.}
\label{figure3}
\end{figure}

As we pointed out in Sec.~\ref{EFT}, the suppression of loops in the perturbative expansion of the partition function depends on the
spatial dimension. For two-dimensional ferromagnetic systems, loops are suppressed by two powers of momentum, in three spatial dimensions,
on the other hand, loops are suppressed by three powers of momentum. Accordingly, the organization of Feynman diagrams related to the
three-dimensional ferromagnet, depicted in Fig.~\ref{figure3} (for details see Ref.~\citep{Hof11a}), is quite different from the one
referring to the two-dimensional ferromagnet, Fig.~\ref{figure1}. Still, as we now discuss, there are also various similarities between
$d_s$=2 and $d_s$=3.

In either case the leading temperature-dependent contribution to the free energy density is related to a one-loop graph. For the
three-dimensional ideal ferromagnet it is of order $p^5 \propto T^{5/2}$, for the two-dimensional ideal ferromagnet we have
$p^4 \propto T^2$. The next-to-leading contribution again stems from a one-loop graph, but this time with an insertion from
${\cal L}^4_{eff}$. For the three-dimensional ferromagnet, this term -- diagram 7 of Fig.~\ref{figure3} -- is of order
$p^7 \propto T^{7/2}$. For the two-dimensional ferromagnet, diagram 6b of Fig.~\ref{figure1} leads to a term of order $p^6 \propto T^3$.

Beyond one-loop order the spin-wave interaction comes into play. However, the corresponding two-loop graph (graph 6a for $d_s$=2, graph 8
for $d_s$=3) which would represent the first candidate for the spin-wave interaction, turns out to be zero due to parity. Regarding the
three-dimensional ideal ferromagnet there is thus no contribution of order $p^8 \propto T^4$ in the series for the free energy density.
Likewise, for the two-dimensional ideal ferromagnet this means that the contribution of order $p^6 \propto T^3$ in the free energy density
is exclusively related to noninteracting spin waves.

So at which order does the spin-wave interaction manifest itself in either case? In three dimensions it shows up at order
$p^{10} \propto T^5$ due to the two-loop diagrams which involve insertions from the next-to-leading Lagrangian ${\cal L}^4_{eff}$, i.e.
graphs 10a and 10b of Fig.~\ref{figure3}. In the case of the two-dimensional ideal ferromagnet the spin-wave interaction sets in at order
$p^8 \propto T^4$ through five different diagrams -- two-loop as well as three-loop diagrams (see Fig.~\ref{figure1}). Note that for the
three-dimensional ferromagnet these three-loop diagrams are of order $p^{11} \propto T^{11/2}$, i.e. of higher order than the two-loop
diagrams 10a and 10b with insertions from ${\cal L}^4_{eff}$. In two spatial dimensions, on the other hand, they are of the same order as
the two-loop diagrams, all five graphs contributing at the same order $p^8 \propto T^4$. The explicit evaluation of these contributions is
quite involved and will be presented elsewhere \citep{Hof12} -- here we rather want to draw our attention to the general structure of the
low-temperature series.

Still, we have to mention that for the two-dimensional ideal ferromagnet, unlike for the ideal ferromagnet in three spatial dimensions,
logarithmic contributions in the low-temperature series will show up. This is related to the structure of the ultraviolet divergences
arising in higher-order loop-diagrams, which in the case of the two-dimensional ferromagnet require a logarithmic renormalization of
next-to-leading order effective coupling constants -- again, details will be provided in Ref.~\citep{Hof12}.

Summarizing the above results, the low-temperature expansion for the free energy density of the ideal ferromagnet in two and three spatial
dimensions -- in the absence of an external magnetic field -- exhibits the following general structure,
\begin{eqnarray}
z_{d_s=2} & = & - {\tilde \eta}_0 T^2 - {\tilde \eta}_1 T^3 + {\cal O}({\bf T^4 \, ln T}, T^4) \, , \nonumber \\
z_{d_s=3} & = & - {\tilde h}_0 T^{\frac{5}{2}} - {\tilde h_1} T^{\frac{7}{2}} - {\tilde h}_2 T^{\frac{9}{2}} - {\bf {\tilde h_3} T^5}
+ {\cal O}({\bf T^{11/2}}) \, ,
\end{eqnarray}
where we have highlighted all terms which are related to the spin-wave interaction.

Note that in the series for the two-dimensional ideal ferromagnet no half-integer powers of the temperature occur. The first contribution
is of order $T^2$ and any other corrections necessarily involve integer powers of the temperature. This is because each additional loop
yields an additional power of $T$. Likewise, higher-order vertices with insertions from the effective Lagrangian,
\begin{equation}
{\cal L}_{eff} = {\cal L}^2_{eff} + {\cal L}^4_{eff} + {\cal L}^6_{eff} + {\cal O}(p^8) \, ,
\end{equation}
also increase the temperature power in steps of $p^2 \propto T$.

Now in three dimensions, the first contribution in the free energy density is of order $T^{5/2}$. Also here, insertions of
higher-order contributions from the effective Lagrangian yield additional integer powers of the temperature. Successive insertions of
higher-order vertices in one-loop graphs, e.g., lead to the pattern $T^{7/2}, T^{9/2}, T^{11/2}, \dots$, describing the effect of
noninteracting spin waves. However, since loops in three spatial dimensions are suppressed by three powers of momentum, or equivalently,
lead to additional powers of $T^{3/2}$, we will also have integer powers of the temperature in the above series. In fact, any such integer
power in the series for the three-dimensional ferromagnet necessarily must have its origin in the spin-wave interaction.

\section{Conclusions}
\label{Summary}

The present study was devoted to the thermodynamic properties of two-dimensional ideal ferromagnets at low temperatures. Previous
articles, based on modified spin-wave theory or Schwinger Boson mean field theory, have also discussed the low-temperature behavior of
two-dimensional ferromagnets, but there the magnons were considered as ideal Bose particles -- the problem of the spin-wave interaction
was neglected and it thus remained unclear whether the low-temperature series given in these articles are complete or will receive
additional corrections due to the interaction. Furthermore, it has never been discussed in a systematic manner how a weak external
magnetic field manifest itself in the low-temperature series for the thermodynamic quantities or how these series look like for underlying
geometries other than a square lattice.

Within the effective Lagrangian framework, we have addressed all these questions in detail. We have derived the low-temperature expansion
of the partition function up to two-loop order -- i.e. order $T^3$ -- for two-dimensional ideal ferromagnets on the square, honeycomb,
triangular and Kagom\'e lattice, where the O(3) spin rotation symmetry is spontaneously broken to O(2) by the ground state. Remarkably,
the spin-wave interaction does not yet manifest itself at order $T^3$ in the free energy density -- it will only enter at order
$T^4 \, \ln T$. Analogously, in the case of the three-dimensional ideal ferromagnet, the spin-wave interaction does not yet manifest
itself at order $T^4$ -- rather, as Dyson showed, it enters at order $T^5$ in the free energy density. In both cases the spin-wave
interaction is thus very weak.

While the validity of the low-temperature series derived within the framework of modified spin-wave theory -- which resorts to an ad hoc
assumption -- has never been critically examined, here we have put these series on safe grounds. Indeed, as discussed in detail, the
Mermin-Wagner theorem is perfectly consistent with our results and the low-temperature series are valid as they stand.

In conclusion, the effective field theory method is a very powerful tool to analyze the general structure of the low-temperature expansion
of the partition function for systems exhibiting collective magnetic behavior. Not only have we conclusively discussed the effect of the
spin-wave interaction and a weak magnetic field in a systematic manner, but also have we put our results on a firm basis. In a more
realistic approach to ferromagnetic films, one should also include the magnetic anisotropy and dipolar interactions. Here, much like
Dyson, Takahashi and other authors, we have considered the ideal ferromagnet and rigorously derived the low-temperature properties for
this "clean" system.

\section*{Acknowledgments}
The author would like to thank H. Leutwyler and U.-J. Wiese for stimulating discussions and for useful comments regarding the manuscript.

\end{document}